# AMIDER: A Multidisciplinary Research Database and Its Application to Promote Open Science

Masayoshi Kozai
https://orcid.org/0000-0002-3948-3666
Polar Environment Data Science Center, Joint Support-Center for Data Science Research, Research Organization of Information and Systems, Tachikawa, Japan
kozai.masayoshi@nipr.ac.jp

Yoshimasa Tanaka
https://orcid.org/0000-0001-7276-4868
Polar Environment Data Science Center, Joint Support-Center for Data Science Research, Research Organization of Information and Systems, Tachikawa, Japan
National Institute of Polar Research, Research Organization of Information and Systems, Tachikawa, Japan

Shuji Abe
https://orcid.org/0000-0001-8270-0024
International Research Center for Space and Planetary Environmental Science, Kyushu University, Fukuoka, Japan

Yasuyuki Minamiyama
https://orcid.org/0000-0002-7280-3342
Research Center for Open Science and Data Platform, National Institute of Informatics, Research Organization of Information and Systems, Tokyo, Japan

Atsuki Shinbori
https://orcid.org/0000-0002-6018-7431
Institute for Space and Earth Environmental Research Center for Integrated Data Science, Nagoya University, Nagoya, Japan

Akira Kadokura
https://orcid.org/0000-0002-6105-9562
Polar Environment Data Science Center, Joint Support-Center for Data Science Research, Research Organization of Information and Systems, Tachikawa, Japan

## Author Contributions

Masayoshi Kozai, Yoshimasa Tanaka, Shuji Abe, Yasuyuki Minamiyama, Atsuki Shinbori, and Akira Kadokura are members of the AMIDER project. They are involved with the developments and activities described in this manuscript. M. K. wrote this manuscript and is conducting system development and operation of the AMIDER. M. K. also performed the text analysis presented in this paper. Y. T. is conducting data collection and curation for the AMIDER database. S. A. developed functional processes used in the AMIDER website. Y. M. is a supervisor from the viewpoints of library information science and the NII RDC project. A. S. is a supervisor from the viewpoint of the IUGONET project. A. K. is a director of PEDSC and supervises its projects, including the AMIDER.

## Abstract

The AMIDER, Advanced Multidisciplinary Integrated-Database for Exploring new Research, is a newly developed research data catalog to demonstrate an advanced database application. AMIDER is characterized as a multidisciplinary database equipped with a user-friendly web application. Its catalog view displays diverse research data at once beyond any limitation of each individual discipline. Some useful functions, such as a selectable data download, data format conversion, and display of data visual information, are also implemented. Further advanced functions, such as visualization of dataset mutual relationship, are also implemented as a preliminary trial. These characteristics and functions are expected to enhance the accessibility to individual research data, even from non-expert users, and be helpful for collaborations among diverse scientific fields beyond individual disciplines. Multidisciplinary data management is also one of AMIDER's uniqueness, where various metadata schemas can be mapped to a uniform metadata table, and standardized and self-describing data formats are adopted. AMIDER website (https://amider.rois.ac.jp/) had been launched in April 2024. As of July 2024, over 15,000 metadata in various research fields of polar science have been registered in the database, and ~500 visitors are viewing the website every day on average. Expansion of the database to further multidisciplinary scientific fields, not only polar science, is planned, and advanced attempts, such as applying Natural Language Processing (NLP) to metadata, have also been considered.

## 1. Introduction

Polar science features diverse scientific fields, including space and upper-atmospheric sciences, meteorology, glaciology, bioscience, geoscience, etc. Sharing the research data in those various fields is essential for polar science because it addresses complex natural systems requiring an inclusive investigation of diverse research fields. In 2017, PEDSC (Polar Environment Data Science Center) was established as one of the centers belonging to the Joint Support-Center for Data Science Research (DS) of the Research Organization of Information and Systems (ROIS), Japan, to accelerate the data-related activities in polar science (Kadokura *et al.*, 2022). NIPR (National Institute of Polar Research), which also belongs to ROIS, and PEDSC have been operating the NIPR Science Database (NIPR, 2024d; Kanao, Okada and Kadokura, 2014; Kanao *et al.*, 2018) and ADS (Arctic Data archive System) (NIPR, 2024b) as a database system for polar science. Research data in polar science have also been published through the Polar Data Journal (NIPR, 2024c; Minamiyama *et al.*, 2017), a data journal of NIPR launched in 2017. NIPR and PEDSC are also key members of the IUGONET (Inter-university Upper atmosphere Global Observation NETwork) project, which provides an integrated metadata catalog and data analysis tools for upper-atmosphere research (IUGONET Project Team, 2024; Tanaka *et al.*, 2023; Yatagai *et al.*, 2014; Hayashi *et al.*, 2013; Tanaka *et al.*, 2013).

Recently, growing trends of open science and multidisciplinary science have raised a new direction emphasizing making the data more open and promoting cross-disciplinary exchanges of researchers and data. We have developed and launched the AMIDER (Advanced Multidisciplinary Integrated-Database for Exploring new Research; https://amider.rois.ac.jp/) on April 23, 2024, to address these directions and develop a next-generation data-sharing platform.

## 2. Methodology

The AMIDER system was designed to overcome the limitations of traditional domain-specific data-sharing platforms by offering a unified catalog of multidisciplinary research data opened to non-expert users. While keeping direct data curation with individual data providers (scientists), which is common in domain-specific databases, AMIDER provides multidisciplinary data browsing. This direct data curation enables the content preparation suitable for user-friendly interfaces described in Section 3; some of them are rarely seen in multidisciplinary data search engines such as CiNii Research (National Institute of Informatics, 2024) or Polar Data Search (Verhey, Minch and Payne, 2023), which are used in Japanese or polar science communities respectively.

To ensure compatibility across scientific fields and interoperability with outside databases, the AMIDER system integrates multiple standard metadata models: the SPASE schema for physics observation data and the ISO/TC 211 schema for geographic information and polar science data. This integration is a compatible concept with the user-friendly interface described above; The AMIDER website displays only basic elements of the metadata, such as contact address and license information, for accessibility from non-expert users. Such an element is common in any metadata schema and can be mapped with each other.

Original data files are managed through a distributed approach where metadata is stored centrally while the data files remain in remote repositories managed by the data providers. Details of these metadata and data management methods are described in Section 4.

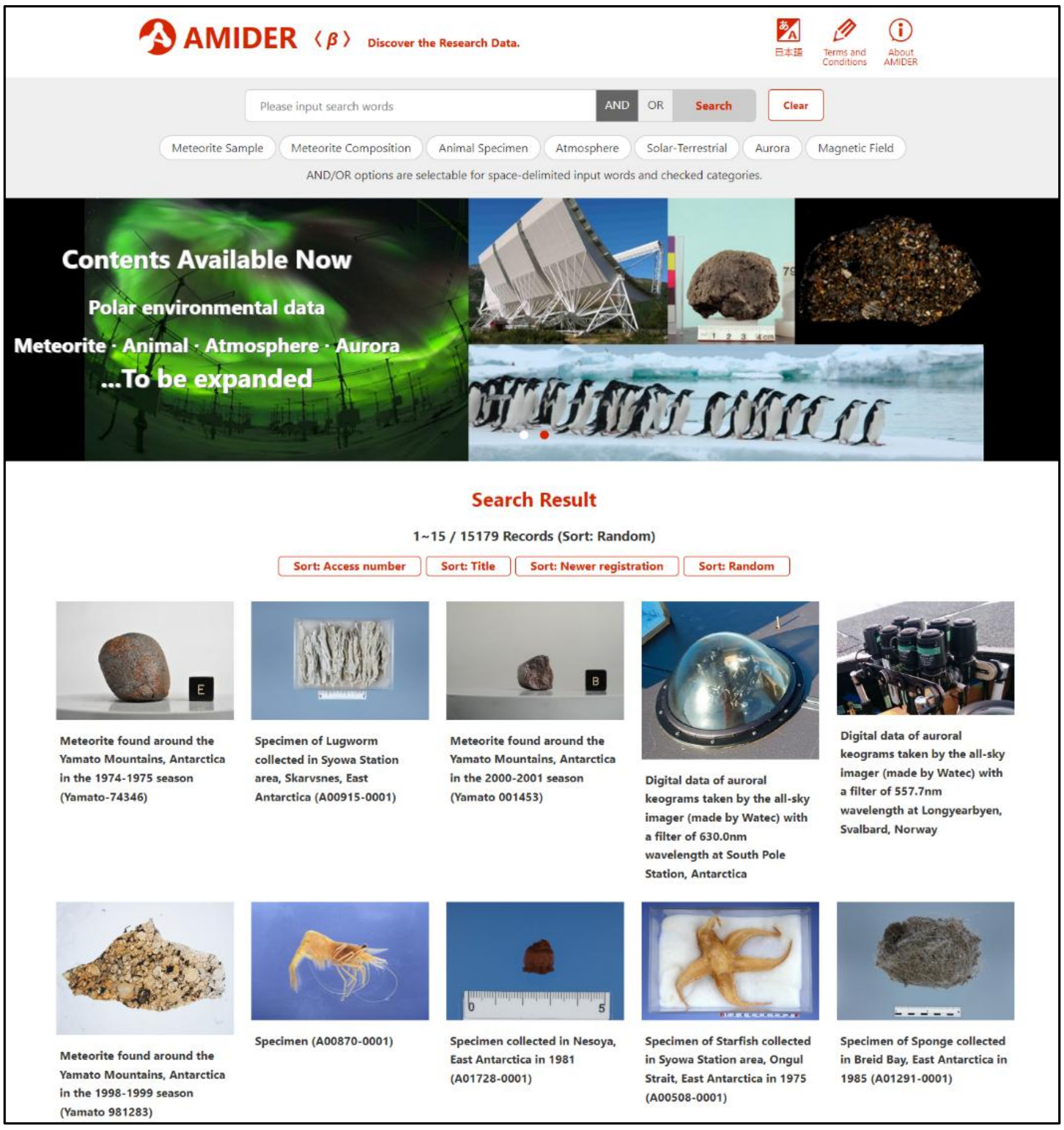

Figure 1: AMIDER's catalog view on the top page (https://amider.rois.ac.jp/).

## 3. Web Application Design

The AMIDER's concept is represented by a catalog view, which offers access to multidisciplinary research data. This catalog view is the front interface users encounter on the AMIDER homepage (https://amider.rois.ac.jp/; Figure 1), providing a comprehensive view of research data across various disciplines. The search result page appears when clicking the "Search" buttons with some search words. It keeps almost the same design as the top page catalog view, providing users with a seamless page transition. The catalog view is inspired by common designs in web marketing, such as electric commerce sites or streaming sites. Each dataset in the catalog view consists of a thumbnail image and snippet, enabling users to grasp individual contents or datasets at a glance. Our data curation process with data providers contributes to forming this catalog view, where impressive images are selected and snippets, or data titles, are written in plain words to help non-expert users. Japanese language is selectable in addition to English, enhancing the accessibility from Japanese users.

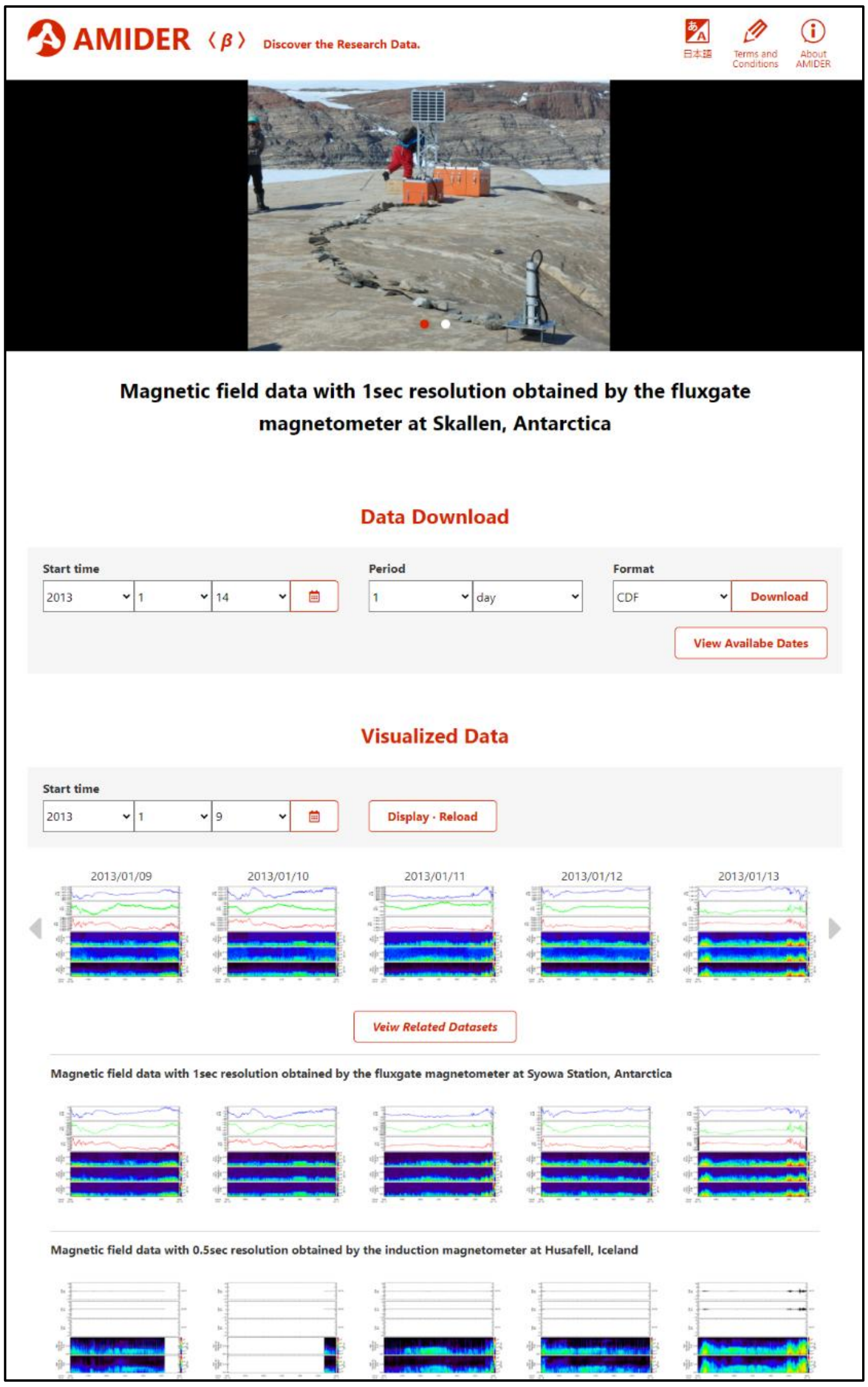

Figure 2: Example of the individual dataset page.

We provide a random sorting function among the sorting options for the catalog view, including the order of access numbers or dataset titles. It randomly extracts datasets from all registered scientific domains, embodying AMIDER's concept of the multidisciplinary database. This sorting option is used on the top page at the first access.

In the search field above the catalog view, some preset keywords such as "Meteorite Sample," "Animal Specimen," and "Aurora" are provided as multi-select chips. AND/OR search options for these categories and space-delimited words input in the text box are selectable by a toggle button.

Figure 2, or https://amider.rois.ac.jp/data/5/, is an example of the individual dataset page for physical observation data. Users can jump to this page only by clicking each thumbnail in the catalog view, and all functions related to each dataset are consolidated on this page. The main visuals and data title at the top of the individual dataset page briefly represent the dataset content. Functions meeting requirements not only from non-expert users but also from scientific specialists are implemented below the title.

For time-series data, the observation period is selectable in the "Data Download" section, and users can download original research data by clicking the "Download" button. Multiple data files are downloadable at once as a zip archive. In the case that the original data is a self-describing binary format in the CDF (NASA/GSFC Space Physics Data Facility, 2024) or NetCDF (UCAR, 2024), which is commonly used in space science or meteorology respectively, AMIDER provides not only its direct download but also its conversion into a plain text (ASCII) format. Another helpful function is the "View Available Dates" button, where users can overview which dates the observation data exists or not in a calendar.

The "Visualized Data" section basically displays data plots for numerical data or specimen photos for specimen-type data. The observation period is selectable for time-series data in the same manner as the data download function. Just below this section, the "Related Datasets" function provides one of our attempts toward a next-generation database application. Correlation scores between datasets are calculated and registered in the database in advance, and datasets having relatively high scores with the dataset on this page are listed in this section. Users can jump to another related dataset by clicking a list item, provided with a walk-around experience between datasets. Currently, only the Pearson correlation coefficient between original numerical data is used as the score for time-series data. The Earth Mover's Distance (EMD) (Rubner, Tomasi and Guibas, 1998) is also used in other numerical data types, such as chemical composition data. The CDF and NetCDF formats are acceptable for calculations of these correlation scores. Further development is planned for this function, including the text mining of metadata to extract the correlations. It will be applicable to any dataset, even if it is specimen-type data without numerical data, and will visualize a cross-disciplinary relationship between datasets.

Information on metadata is also displayed at the bottom of this page (not shown in Figure 2). Only selected elements of original metadata that are necessary and sufficient for the basic use of the data are displayed here, prioritizing accessibility from non-expert users. In specimen-type data, the metadata table is essential for users to find and contact specimen owners to observe or send a request for the real specimens.

## 4. System and Data Management

This section describes AMIDER's backend system and data management, which realize the user interfaces described in Section 3. Figure 3 overviews the AMIDER system and its data flow. AMIDER's core content is the metadata catalog. Therefore, the metadata file is the minimum required data to register research data into the AMIDER system. Two XML schemas are acceptable as the metadata format: the SPASE schema (Roberts *et al.*, 2018; SPASE Consortium, 2024) and the ISO/TC 211 Geographic information/Geomatics schema (ISO/TC 211, 2024). These standardized schemas allow the AMIDER to use the XML framework and ensure interoperability and collaboration with outside databases via the metadata. The SPASE schema is developed as a standard metadata model for observation data in space and upper-atmospheric sciences. It suits physics observation data and ensures the AMIDER's interoperability with other space and upper-atmospheric science databases such as IUGONET. The ISO schema is one of the schemas used in the ADS, an official research data repository of NIPR and JARE (Japanese Antarctic Research Expedition) project. Therefore, accepting this schema enables the AMIDER to collaborate with the polar science community in Japan.

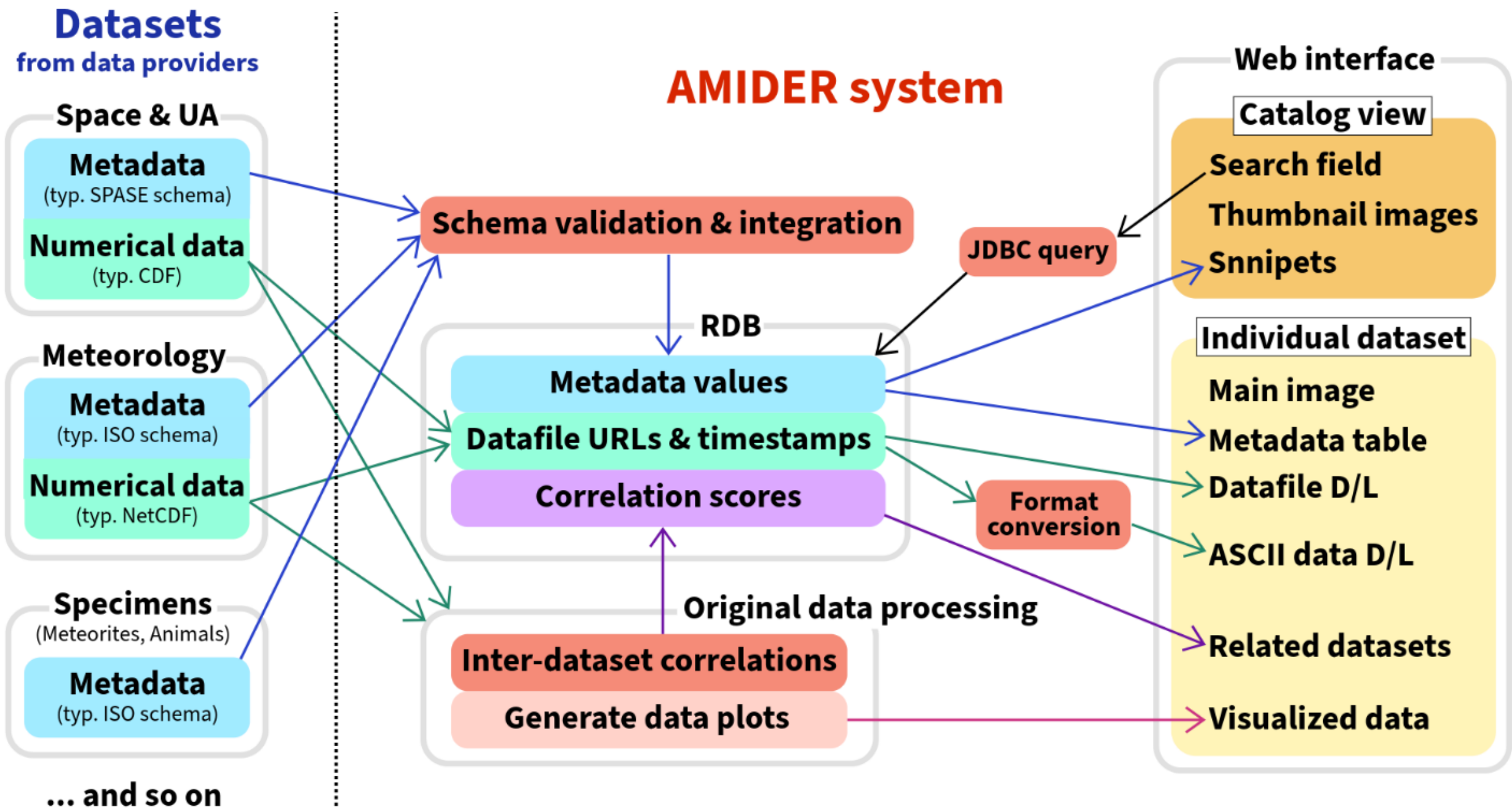

Figure 3: AMIDER system and data flow. Metadata of different-discipline data, such as space and upper-atmospheric (UA) science and meteorological data, are validated for each XML schema and integrated into AMIDER's unified database. Original numerical data are also processed via their URLs for AMIDER's functions such as "Related Datasets."

The submitted XML file is validated for the corresponding schema, integrated into AMIDER's unified namespace, and then registered into the relational database (RDB), which forms the main database system of AMIDER. PostgreSQL (The PostgreSQL Global Development Group, 2024) is used for the RDB management system (RDBMS). The search function in the catalog view queries for search keywords input from the search field, in the RDB via the Java Database Connectivity (JDBC).

For the original numerical data, such as observation data, AMIDER's database stores only its URLs in remote repositories managed by each data provider. This policy enables the centralized management of datafiles in the original repository. The "Data Download" function in the individual dataset page accepts any format of datafiles as long as the URLs are provided, while other advanced functions, including the ASCII conversion and correlation score calculation, are provided for the CDF and NetCDF as described in Section 3. These original data processing are one of AMIDER's uniqueness beyond a mere metadata catalog.

Image files used in the catalog view and individual dataset page (main visual and "Visualized Data" section) are provided in the PNG or JPG format and stored in the AMIDER server. As one of the original data processing functions, AMIDER can also offer a data plot creation from the CDF or NetCDF data URLs. For the time-series data, the corresponding time of each data plot is indicated by its file name. A configuration file in YAML format defines the naming rule, such as "%YYYY-%mm-%dd". Other display options, such as displaying or not-displaying the visualized data section, are also selectable in the configuration file for each dataset. This configuration file is prepared by each data provider, enabling customization of each dataset page while keeping a uniform web design in AMIDER.

## 5. Results

As of July 2024, the AMIDER database has accumulated more than 15,000 metadata entries mainly from polar science researchers in Japan, encompassing datasets related to space and upper-atmospheric observations, meteorological data, meteorite collections, and biological specimens from polar regions. The data publication activities, such as IUGONET, in space and upper-atmospheric science are mainly aiming at activating data analysis studies in solar-terrestrial physics. AMIDER is more focused on data promotion, providing a complementary role to the conventional databases. NIPR's meteorite collection features one of the largest collections of extraterrestrial material in the world (NIPR, 2024a). There is also a domain-specific database for meteorites (The Meteoritical Society, 2024), while AMIDER's meteorite contents will enhance their visibility to non-expert users. AMIDER also offers a publication site for animal specimens collected by JARE, for which there is no other public database currently. The meteorological datasets contain field observation data by JARE, such as carbon dioxide concentration in the Antarctic atmosphere, and will provide a unique contribution to climate change research. Efforts to broaden the contents to a wider range of scientific disciplines are currently in progress, with the goal of enhancing multidisciplinary data exploration. Since AMIDER's launch in April 2024, the system has averaged approximately 500 unique visitors daily, demonstrating substantial interest from the research community.

## 6. Future Prospects

We are considering further advanced attempts to contribute to open science by utilizing the AMIDER system as a demonstration field in parallel with its website's operation. This section briefs our future prospects including such attempts.

Beyond a mere metadatabase that only displays datasets, actively inducing cross-disciplinary user access to each content will be the next issue of the database applications. AMIDER's "Related Datasets" function is an attempt at such an issue, and this concept has more room for advanced development. Figure 4 shows a co-occurrence network extracted from titles of space and upper-atmospheric datasets in the AMIDER database, created to provide an advanced visualization of the research data. Each circle with a label indicates a noun word extracted by morphological analysis of the dataset titles. The circle size is proportional to the occurrence rate of each word, and lines express the co-occurrence of the words in each title. It visualizes that diverse observation data in this field form a network via some terms, such as an observation site "Syowa Station," scientific target "cosmic noise absorption," or an experimental device "magnetometer." These extracted terms and their network will allow non-expert users to grasp the diverse research data at a glance and explore the data following their interests. We are now attempting multidisciplinary text mining in more diverse scientific fields in the AMIDER metadatabase. These attempts will also contribute to promoting multidisciplinary research in the future.

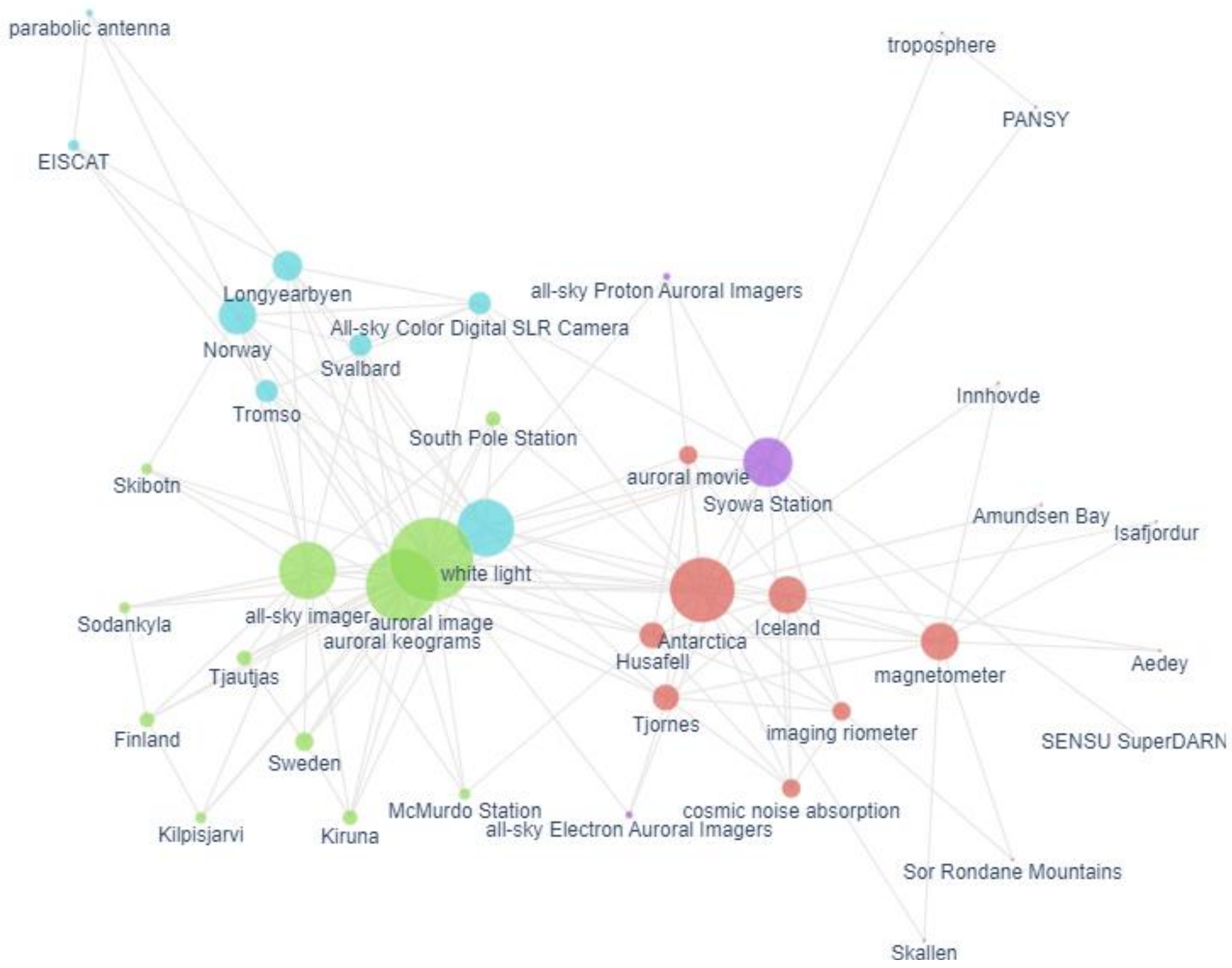

Figure 4: Co-occurrence network of titles for space and upper-atmospheric datasets in the AMIDER database.

Text mining or natural language processing (NLP) is expected to be applied not only to the data visualization mentioned above but also to other developments related to text data, such as a metadata creation tool or interoperation with outside databases via metadata. Aiming at these developments, we have started collaborations with the Research Data Cloud (RDC) project of NII, Japan, through its use-case creation project (MEXT, 2024), and with researchers of the NLP application (JSPS KAKENHI, 2024).

## 7. Summary

Sharing research data, such as data obtained in scientific experiments, forms a basis for studying complex systems in nature. There are well-established data-sharing platforms in individual scientific disciplines, but a multidisciplinary platform is desired to make the research data more open. Research data are direct outputs of research activities, and their visualization by an advanced data catalog will further activate the scientific fields.

A novel research data catalog, AMIDER, has been developed to address these requirements and launched in April 2024. It combines a multidisciplinary database and a user-friendly web application, and the target users are non-expert users such as researchers interested in multidisciplinary research, educators, or students. Its catalog view, consisting of thumbnails and snippets, allows users to grasp the diverse datasets at a glance. Functions in the individual dataset page, such as the “Data Download” and “Visualized Data,” meet requirements from not only non-expert users but also individual specialists. The “Related Datasets” function is implemented as one of the attempts for the next-generation database; it proposes relationships between the datasets and is expected to accelerate each user’s data exploration.

The domain-specific databases have been enhancing their functionality, thanks to the uniform format and established usage method of their data. On the other hand, AMIDER’s challenge is to achieve user-friendly functions in the multidisciplinary database, aiming at bridging the data providers (scientists) and non-expert users. Especially the web-marketing-inspired and multidisciplinary catalog view, ASCII conversion function from binary data formats, and “Related Datasets” function are rarely seen in other research databases, enhancing the uniqueness of the AMIDER. We have started further advanced attempts for the promotion of research data, such as leveraging the text mining technique for data visualization. Their implementations in the AMIDER system in the near future will provide a demonstration of the next-generation system.

## Competing Interests

The authors have no competing interests to declare.

## Acknowledgments

This work was supported by MEXT as “Developing a Research Data Ecosystem for the Promotion of Data-Driven Science” and by JSPS KAKENHI Grant Number JP24K14966.